\documentclass[11pt]{article}
\usepackage{moriond,epsfig}
\usepackage[latin1]{inputenc}
\usepackage[OT1]{fontenc}

\def\Journal#1#2#3#4{{#1} {\bf #2}, #3 (#4)}

\def\apj{\em ApJ}
\def\apjl{\em ApJL}
\def\apjs{\em ApJS}
\def\aj{\em AJ}
\def\mnras{\em MNRAS}

\def\be{\begin{equation}}
\def\ee{\end{equation}}
\def\bea{\begin{eqnarray}}
\def\eea{\end{eqnarray}}

\def\hmpc{$h^{-1}$Mpc }
\def\gtwid{\mathrel{\raise.3ex\hbox{$>$\kern-.75em\lower1ex\hbox{$\sim$}}}}
\def\ltwid{\mathrel{\raise.3ex\hbox{$<$\kern-.75em\lower1ex\hbox{$\sim$}}}}

\def\etal{{\it et al.\ }}
\begin{document}
%\vspace*{4cm}
\title{Cosmic Flows on 100 $h^{-1}$Mpc scales}

\author{Hume A. Feldman$^{\star,1}$, Michael J. Hudson$^{\dagger,2}$  \& Richard Watkins$^{\ddagger,3}$ }

\address{$^\star$Department of Physics \& Astronomy, University of Kansas, Lawrence, KS 66045, USA.\\
$^\dagger$Department of Physics and Astronomy, University of Waterloo, Waterloo, ON N2L 3G1, Canada.\\
$^\ddagger$Department of Physics, Willamette University, Salem, OR 97301, USA.\\
emails: $^1$feldman@ku.edu;\, $^2$mjhudson@uwaterloo.ca;\, $^3$rwatkins@willamette.edu}

\maketitle\abstracts{
To study galactic motions on the largest available scales, we require bulk flow moments whose window functions have as narrow a peak as possible and having as small an amplitude as possible outside the peak. Typically the moments found using the maximum likelihood estimate weights do not meet these criteria. We present a new method for calculating weights for moments that essentially allow us to "design" the moment's window function, subject, of course, to the distribution and uncertainties of the available data.    }

On scales that are small compared to the Hubble radius, 
galaxy motions are manifest in deviations from the idealized 
isotropic cosmological expansion
\be
cz=H_0r + \hat{\bf r}\cdot\left[{\bf v}({\bf r})-{\bf v}({0})\right]\ .
\ee
Redshift-distance samples, obtained from peculiar velocity 
surveys, allow us to determine the radial (line-of-sight) 
component of the peculiar velocity of each galaxy:
\be
v(r)=\hat{\bf r}\cdot{\bf v}({\bf r})=cz-H_0r\ .
\ee

Galaxies trace the large-scale linear velocity field $v(r)$ which is 
described by a Gaussian random field that is completely defined, 
in Fourier space, by its velocity power spectrum $P_v(k)$. 
The Fourier Transform of the line-of-sight velocity is
\be
\hat{\bf r}\cdot{\bf v}({\bf r})=\frac1{(2\pi)^3}\int\ d^3{\bf k}\ \hat{\bf r}\cdot\hat{\bf k}\ v({\bf k}){\rm e}^{i{\bf k}\cdot{\bf r}}\ ,
\ee
which provides us with the velocity power spectrum
\be
<v({\bf k})v^*({\bf k^\prime})>=(2\pi)^3P_v(k)\delta_D({\bf k}-{\bf k}^\prime)\ ,
\ee
where $\delta_D$ is the Dirac delta function. In linear theory, the velocity power spectrum is related to 
the density power spectrum 
\be 
P_v(k)=\frac{H^2}{k^2}f^2(\Omega_{m,0}\Omega_\Lambda)P(k)\ ,
\ee
where $H$ is the Hubble constant and $f$ is the rate of perturbation growth at the present. $\Omega_m$ and $\Omega_\Lambda$ are the matter and dark energy density parameters today.
The power spectrum provides a complete statistical description of the linear peculiar velocity field.

For our power spectrum model, we follow Eisenstein and Hu (1998) in writing 
\be
P(k)\propto\sigma_8\Omega_m^{0.6}k^nT^2(k/\Gamma)\ ,
\ee
where $T$ is the transfer function, $\Gamma$ is the power spectrum ``shape" parameter and as usual, we normalize the power spectrum using the parameter $\sigma_8$, the amplitude of matter density perturbations on the scale of $8\ h^{-1}$Mpc. 

To find the maximum likelihood likelihood parameters from peculiar velocity surveys we start with catalog of peculiar velocities galaxies, labeled by an 
index $n$, positions ${\bf r}_n$, estimates of the line-of-sight peculiar velocities $S_n$ and uncertainties $\sigma_n$. We assume that the observational errors are Gaussian 
distributed. We then model the velocity field as  a uniform streaming motion, or 
bulk flow (BF),  denoted by ${\bf U}$, about which there are random motions 
drawn from a Gaussian distribution with a 1-D velocity dispersion $\sigma_*$.
Then the likelihood function for the BF components is
\be
{\cal L}(U_i) =  \prod_n\frac1{\sqrt{\sigma_n^2+\sigma_*^2}}\exp\left(-\frac12\frac{(S_n-\hat r_{n,i}U_i)^2}{\sigma_n^2+\sigma_*^2}\right).
\ee
where $i$ go from 1 to 3 to specify the BF components.    While this has been done previously for maximum likelihood estimates (MLE) of the bulk flow moments, the advantage of these new moments is that they have been designed to be sensitive {\it only} to scales of order 100$h^{-1}$Mpc.    Thus we will be able to probe these scales without having to worry about the influence of smaller scales.    Further, by isolating the very large scale motions we will be able to put stronger constraints on  power spectrum parameters.    

Maximum likelihood solution for BF moments is
\be
U_i=A_{ij}^{-1}\sum_n\frac{\hat r_{n,j}S_n}{\sigma_n^2+\sigma_*^2}\ ,
\ee
where
\be
A_{ij}=\sum_n\frac{\hat r_{n,i}\hat r_{n,j}}{\sigma_n^2+\sigma_*^2}\ .
\ee
The measured peculiar velocity of galaxy $n$
\be
S_n=\hat r_{n,i}v_i({\bf r}_n)+\epsilon_n\ ,
\ee
where $\epsilon_n$ is a Gaussian with zero mean and variance $\sigma_n^2+\sigma_*^2$.
The theoretical covariance matrix for the BF components 
\be
 R_{ab} = R^{(v)}_{ab} +  R^{(\epsilon)}_{ab} . 
\ee
The first term is given as an integral over the matter fluctuation power spectrum, $P(k)$, 
 \be
 R^{(v)}_{ab}  = {\Omega_{m}^{1.2}\over 2\pi^2}\int_0^\infty
 dk \ \  {\cal W}^2_{ab}(k)P(k),
 \ee
 where the angle-averaged tensor window function is 
 \be
 {\cal W}^2_{ab} (k)= \sum_{n,m} w_{a,n} w_{b,m}\int {d^2{\hat k}\over 4\pi}\ \left({\bf \hat r}_n\cdot {\bf \hat k}\ \ {\bf \hat r}_m\cdot {\bf \hat k}\right)
 \exp\left(i{\bf k}\cdot ({\bf r}_n- {\bf r}_m)\right).
\ee
In order to do better than the MLE solutions, we redesign the window functions by requiring that they have narrow peaks and small amplitude outside the peak.
We start with the BF MLE weights
\be
w_{i,n}=A_{ij}^{-1}\sum_n\frac{{\bf x}_j\cdot{\bf r}_n}{\sigma_n^2+\sigma_*^2}\ ,
\ee
which depends on the spatial distribution and errors. Now consider an ideal survey with very large number of points, isotropic distribution 
and a Gaussian falloff $n(r)\propto\exp(-r^2/2R_I^2)$ where $R_I$ is the depth of the survey.

The moments are specified by the weights 
\be
u_i=\sum_nw_{i,n}S_n
\ee
that minimize the variance $<(u_i-U_i)^2>$. We expand the variance
\be
<(u_i-U_i)^2>=\sum_{n,m}w_{i,n}w_{i,m}<S_nS_m>+<U_i^2>-2\sum_nw_{i,n}<U_iS_n>\ ,
\ee
since the measurement error included in $S_n$ is uncorrelated with the bulk flow $U_i$. 
For BF moments we can impose the constraint that
\be
\lim_{k\to 0} {\cal W}^2_{ii} (k)=   \sum_{n,m} w_{i,n} w_{i,m} \int {d^2{\hat k}\over 4\pi}\ \left({\bf \hat r}_n\cdot {\bf \hat k}\ \ {\bf \hat r}_m\cdot {\bf \hat k}\right)=1/3\ .
\ee

We now minimize this expression with respect to $w_{i,n}$ subject to the constraint which we enforce using a Lagrange multiplier to get
\be
\sum_m \left(\langle S_nS_m\rangle+\lambda P_{nm}\right)w_{i,m}  = \langle S_nU_i\rangle. 
\ee
We solve this to get the minimum variance (MV) weights.

\begin{figure}[t]
\psfig{figure=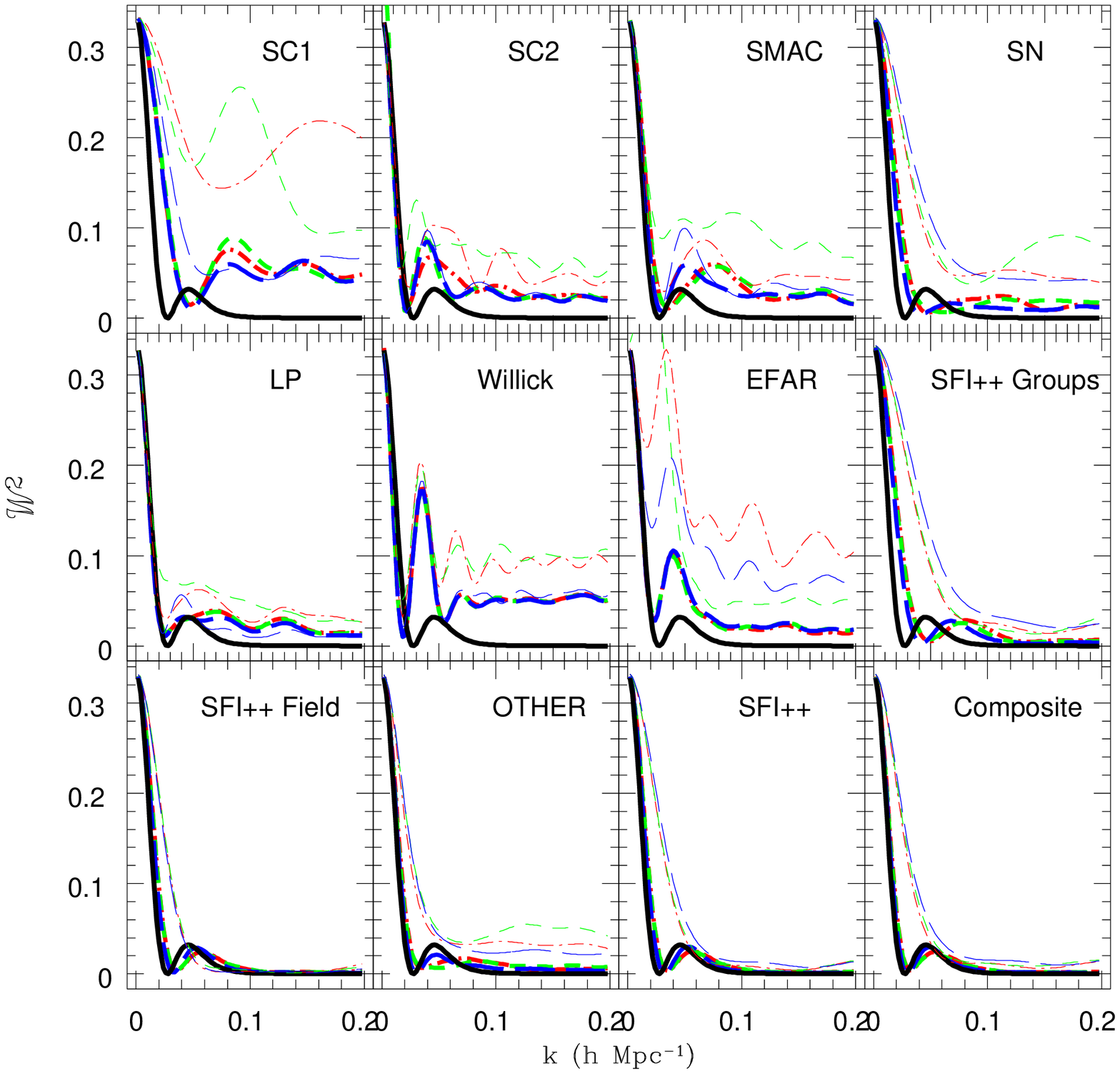,height=8cm}        
\psfig{figure=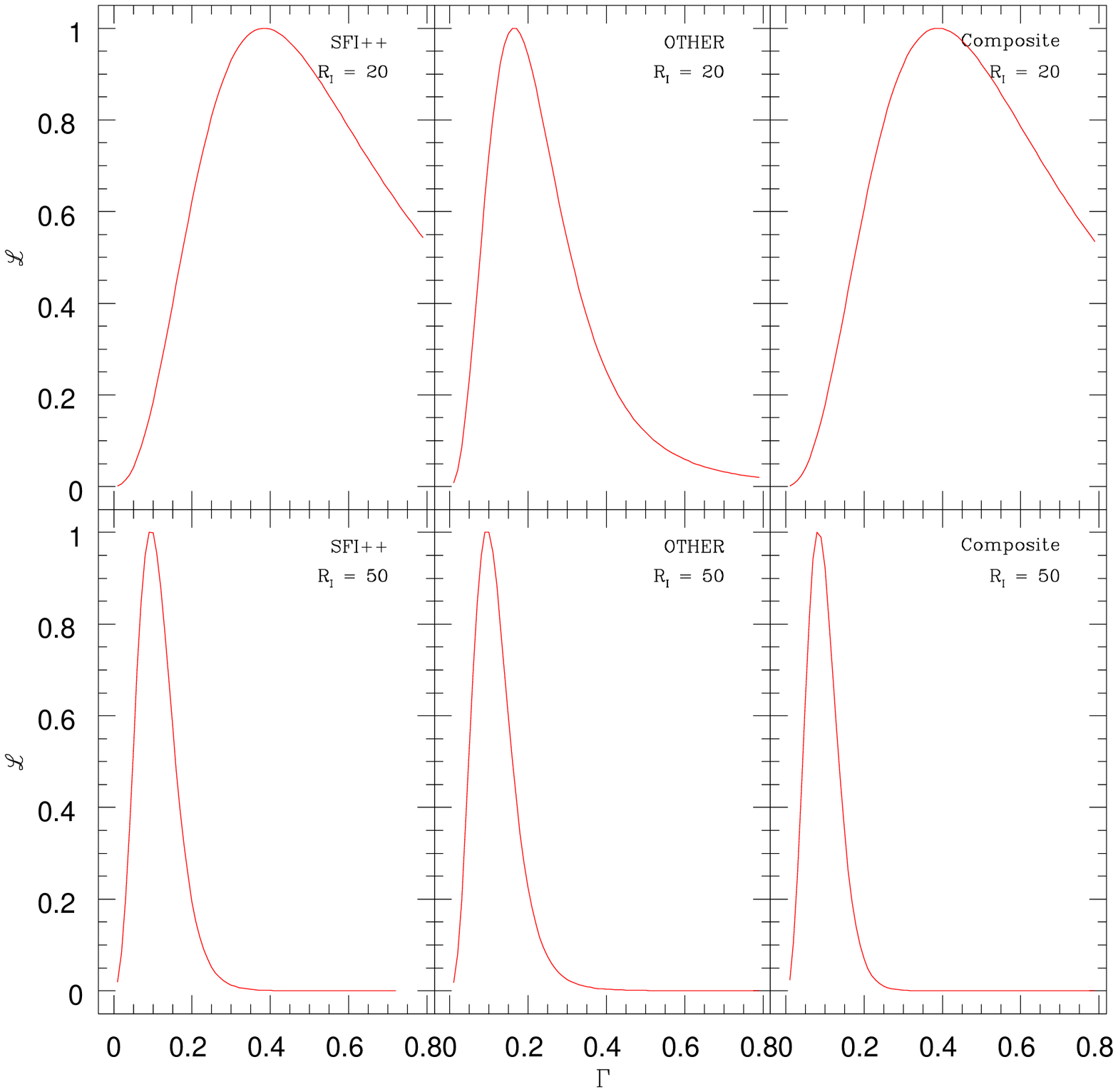,height=8cm}
\parbox{1in}{} \ \hspace{0.2cm} \
\parbox{2.8in}{\scriptsize {\bf Fig. 1:} The window functions of the bulk flow component for $R_I=50$\hmpc\  for the all the catalogs we considered. The thick (thin) lines are the window functions for the MV (MLE) bulk flow components. the x,y,z--component are dash-dot, short dash, long dash lines respectively. The Thick solid line is the ideal window function.}\ \hspace{0.6cm} \
\parbox{2.8in}{\scriptsize {\bf Fig. 2:} The $\Gamma$ likelihood function for the composite catalogs.}
\end{figure}

To apply this formalism, we used both cluster and galaxy peculiar velocity surveys. Specifically, we used the 
Willick cluster survey\cite{willick}, the SC1 \cite{sc1} and SC2 \cite{sc2} (Tully-Fisher clusters);
The EFAR survey \cite{EFAR} (a Fundamental Plane cluster survey);  the SN are Type Ia supernovae from the compilation of  Tonry \etal\cite{TonSchBar03};
SMAC \cite{SMAC1,SMAC2} (a Fundamental Plane cluster survey); SFI++\cite{SFI1,SFI2} (a Tully-Fisher based survey of $\sim2700$ field galaxies and $\sim730$ groups).

The minimum variance, or MV, weights were calculated for the bulk flow component moments using the method described above for each of our catalogs, as well as for three composite catalogs, SFI++, consisting of both fields and groups from the catalog,  OTHER, consisting of all the surveys except for the SFI++ field galaxy and groups  catalogs, and Composite, consisting of all of the catalogs combined.   The OTHER catalog was included due to the fact that, given their large size, the SFI++ catalogs tend to dominate any composite catalog in which they are included.   Here we will show results from a deep survey, $R_I= 50\ h^{-1}$Mpc.

In Figure 1 we show the window functions of the MV bulk flow component moments.   For comparison, we also include the ideal window functions as well as those for the MLE moments for each survey.     As expected, the match between the window functions for the MV moments and the ideal is best for the large surveys and those with small measurement error and similar distribution to the ideal survey.   For the sparse, noisy surveys, the window functions for the MV moments are not very different than those of the MLE moments, differing mostly in the amplitude of the tail of the window function for large $k$.    
   
In Figure 2. we show the likelihood function of the ``shape" parameter $\Gamma$ for the large catalogs. We see clearly that $\Gamma$ is smaller for large $R_I$ than for smaller $R_I$. In Figure 3. we show the BF velocity for the composite surveys. We see that for the shallow window functions, we get small flows, $\sim150$ km/s. On large scales we get very large flows $\sim400$ km/s. This unexpected flow which is exhibited in all the catalogs we analyzed, require a very steep power spectrum which leads to small $\Gamma$ (see figure 2.)

\begin{figure}[t]
\psfig{figure=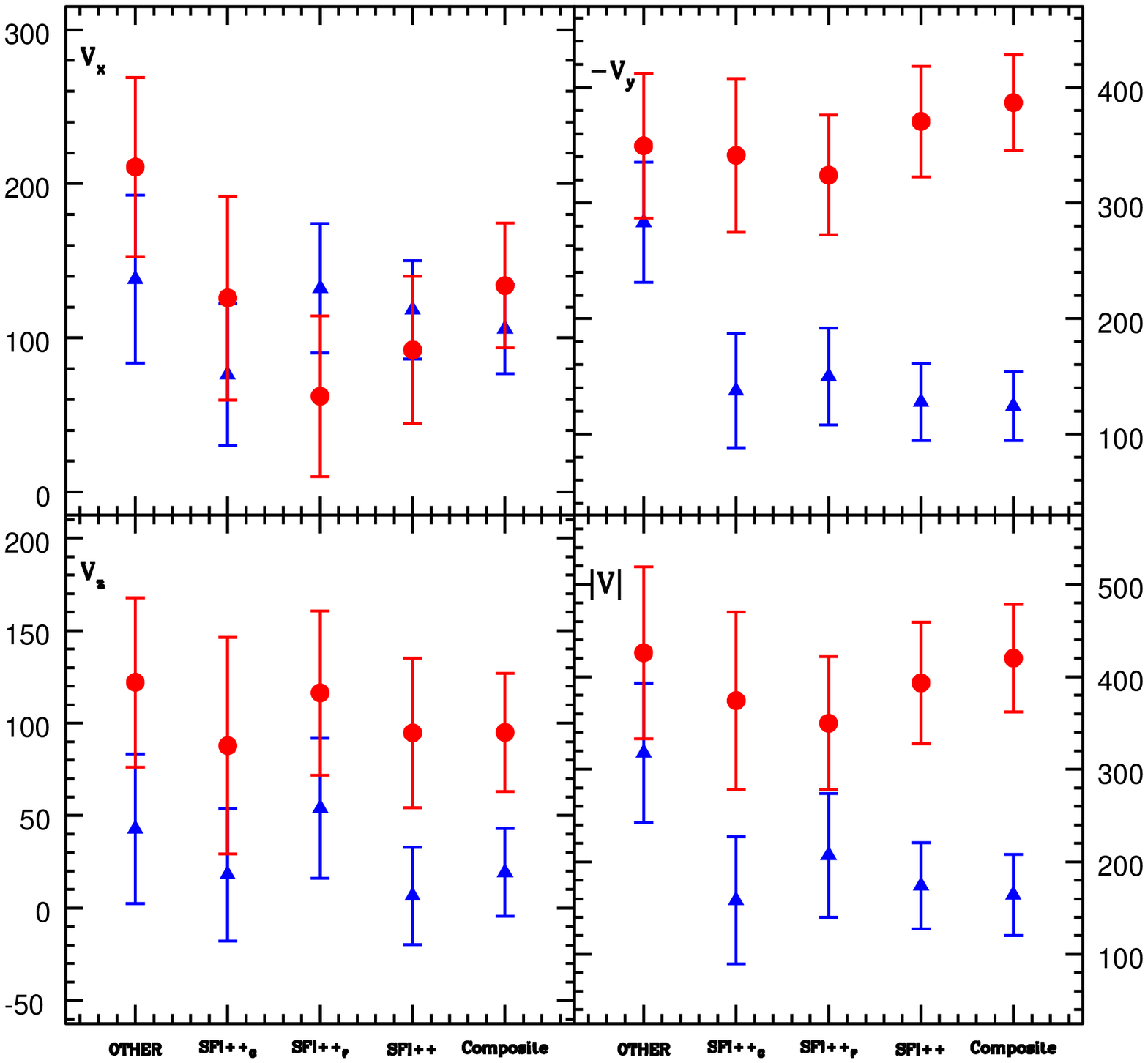,height=10cm}        

\parbox{5in}{\scriptsize {\bf Fig. 3:} The bulk flow of the large catalogs. The triangles (points) are the BF moments with $R_I=20$\hmpc ($R_I=50$\hmpc). There is a consistent and robust flow exhibited in all catalogs to the negative Galactic y direction (upper right panel) for the large scale analysis which is reflected in the BF magnitude (lower right panel)}
\end{figure}

\end{document}